\documentclass[11pt]{article}
\usepackage{amsmath,amsthm,latexsym,amssymb,amsfonts,epsfig}
\usepackage{float}
\usepackage{cite}
\usepackage{hyperref}
\hypersetup{
    colorlinks,%
    citecolor=blue,%
    filecolor=blue,%
    linkcolor=blue,%
    urlcolor=blue
}

\newcommand{\be}{\begin{equation}}
\newcommand{\ee}{\end{equation}}
\newcommand{\ba}{\begin{eqnarray}}
\newcommand{\ea}{\end{eqnarray}}

\title{{\sf Hamiltonian Renormalisation V: Free Vector Bosons}}
\author{
{\sf K. Liegener}$^1$\thanks{{\sf 
klaus.liegener@desy.de}},
{\sf T. Thiemann}$^2$\thanks{{\sf 
thomas.thiemann@gravity.fau.de}}\\
\\
{\sf $^1$ II. Institute for Theoretical Physics, University of Hamburg,}\\
{\sf Luruper Chaussee 149, 22761 Hamburg, Germany}\\
{\sf $^2$ Institute for Quantum Gravity, FAU Erlangen -- N\"urnberg,}\\
{\sf Staudtstrasse 7, 91058 Erlangen, Germany}\\
}
\date{{\small\sf \today}}

\makeatletter
\@addtoreset{equation}{section}
\makeatother

\newcommand{\nocontentsline}[3]{}
\newcommand{\tocless}[2]{\bgroup\let\addcontentsline=\nocontentsline#1{#2}\egroup}

\begin{document}

\maketitle
{\sf
\begin{abstract}
\fontfamily{lmss}\selectfont{
In a recent proposal we applied methods from constructive QFT to derive a Hamiltonian Renormalisation Group in order to employ it ultimately for canonical quantum gravity. The proposal was successfully tested for free scalar fields and thus a natural next step is to test it for free gauge theories. This can be done in the framework of reduced phase space quantisation which allows using techniques developed earlier for scalar field theories. In addition, in canonical quantum gravity one works in representations that support holonomy operators which are ill defined in the Fock representation of say Maxwell or Proca theory. Thus, we consider toy models that have both features, i.e. which employ Fock representations in which holonomy operators are well-defined.  We adapt the coarse graining maps considered for scalar fields to those theories for free vector bosons. It turns out that the corresponding fixed pointed theories can be found analytically.
}
\end{abstract}

\tableofcontents

\newpage

\section{Introduction}
\label{s1}
\numberwithin{equation}{section}
The construction of interacting four-dimensional Quantum Field Theories (QFTs) is an interesting and fundamentally important problem in modern physics. Despite several attempts it has not been satisfactorily completed as of today\cite{WG64,OS73_1,OS75_2,GJ87,Fr78,Riv00,JW00}.
Due to several challenges along the way, preliminary computations are often done in the presence of finite infrared and ultraviolet cut-offs, most prominently in the framework of Lattice Gauge Theories (LGT) \cite{Cre83,HLS17}. Especially considering approaches towards Quantum Gravity, it motivated proposals where the discretisation of space(-time) was assumed to be fundamental \cite{Loll98,AQG1,BD09_a,BD09_b,DRS12,Loll19}. This allowed to make a wide range of predictions by performing computations using established tools from LGT, see for example \cite{BKR17,DL17,GS19,HL20}.\\
However, as it is not yet experimentally supported whether these discrete structures are fundamental, one can independently ask if they can be understood as coarse graining of some underlying continuum QFT and -- of course -- the construction of such a QFT is in itself an aspirational goal. A possible avenue for this comes in the form of {\it inductive limits} \cite{KR86,Jan88,Sau98,Thi07}. This presents a construction by which a QFT described by a Hilbert space $\mathcal{H}$ supporting a Hamiltonian operator $\hat{H}$ can in principle be obtained from a consistent family of discretised theories described by a Hilbert space $\mathcal{H}_M$ supporting a Hamiltonian $\hat{H}_M$ where $M$ labels the different discretisation scales. The necessary condition for the existence of such an inductive limit is that there exists a family of isometric injection maps $J_{M\to M'} : \mathcal{H}_M \to \mathcal{H}_{M'}$   for $M<M'$ in the sense of $M'$ describing finer resolution than $M$. $J_{M\to M'}$ must be subject to a certain {\it compatibility condition} in order to enable the reconstruction of the inductive limit Hilbert space $\mathcal{H}$ and to allow an interpretation of the $\mathcal{H}_M$ as restrictions of $\mathcal{H}$ to coarse resolution $M$. Similarly, there exists a condition for a family of quadratic forms $\hat{H}_M$ which guarantees the existence of a corresponding limit quadratic form  $\hat{H}$ on $\mathcal{H}$.\\
In a recent series of paper \cite{LLT1,LLT2,LLT3,LLT4} we introduced a Hamiltonian formulation of the renormalisation group which is rather close in methodology to density matrix renormalisation \cite{BS19,BS19_b,SMMT20} and projective renormalisation \cite{Yam85,Oko13,KO17,Lan_1,Lan_2,Lan_3,Lan_4,Lan_5} which in turn are based on the seminal ideas of Wilson, Kadanov and Fisher \cite{Wil75,Kad77,Fish74}. The proposal is motivated by formulations of the renormalisation group in the covariant setting \cite{WK73,Wil75,Kad77,Fish74,Has98} which can be reformulated in Hamiltonian terms using Osterwalder-Schrader reconstruction and in fact gives rise to a natural flow of inductive structures and Hamiltonian quadratic forms \cite{LLT1,LLT2}. That the direct {\it Hamiltonian Renormalisation Group} delivers the correct results has been demonstrated for the case of the massive, free scalar field in arbitrary dimensions \cite{LLT2,LLT3,LLT4}. The next challenge for this programme is its extension to gauge theories, as the most interesting models of modern physics are phrased in this language, e.g. QCD. In this paper, we perform the firsts steps in this direction by considering a toy model which is a certain deformation of the reduced Hamiltonian of Maxwell theory. The deformation consists in adding a Proca like mass term to higher powers of the Laplacian in order that the Fock space defined by that Hamiltonian supports holonomy operators, which are exponentials of the connection smeared along one-dimensional curves. The motivation for considering such theories comes from an approach to canonical quantum gravity \cite{Thi07,Rov04} for which holonomies play a fundamental role and are promoted to well defined operators upon quantisation.\\
A possible way to proceed is as follows: prior to quantisation one can transcend to the reduced phase space, where the Gauss constraints have been implemented. Since the gauge-invariant (transversal) modes can be treated as scalars, the tools from \cite{LLT1,LLT2,LLT3,LLT4} become applicable. With them, it is possible to analytically determine the fixed points which lead to the correct continuum theory.\\
Another approach is to implement the Gauss constraint after quantization. This involves adapting 
the coarse graining maps for scalar fields to vector bosons. In particular, this involves smearing 
the field against form factors rather than scalar smearing functions. In this paper we will incorporate 
the latter feature by considering a modification of Proca theory that allows for holonomy operators. The actual solution of the Gauss constraint after quantisation combined with coarse graining will be  subject of a subsequent  paper \cite{tbp}. We will introduce the necessary coarse-graining maps for this procedure and present explicitly how fixed points can be computed in the new setting.\\

The architecture of the article is as follows:\\
In section \ref{s2_ver0} we follow the route of reduced phase space quantisation. The first subsection \ref{s21} reviews the framework of our version of the Hamiltonian Renormalisation Group for scalar fields to familiarise with the notation of this paper and to enable comparison with \cite{LLT1}. We start by first looking at ``classical'' discretisations and define injection and evaluation maps between theories of different resolution. These discretisations are built, e.g., with respect to cuboidal tessellations of our spatial manifold. The second subsection \ref{s22} introduces a ${\rm U}(1)$ toy model with Gauss constraint $G(x)$, which is highly inspired by free Maxwell electrodynamics. 
As an alternative to implementing the Gauss constraint classically, one may introduce a Master Constraint of the form ${\bf M}=\frac{1}{2}\int d^3x\; K(x,x')G(x) G(x)$ and promote it to an operator on the Fock space with some positive kernel $K(x,x')$ analogously to \cite{DT04}. Determining the physical Hilbert space will reduce to the space of transversal modes. This is equivalent to first fixing the gauge on the classical level and then performing a reduced phase space quantisation of the transversal modes. As both methods lead to the same result, we will employ here the latter strategy. In the third subsection \ref{s23} we briefly recall how the tools from \cite{LLT1,LLT2,LLT3,LLT4} find application and lead to the correct fixed points.\\
In section \ref{s3_ver1} we go further into the direction of LGT: we are interested in the connection integrated along edges of the discretising lattices. To bring this formulation close to \cite{LLT1}, in subsection \ref{s31} we define the discretised fields as the continuum fields smeared against (distributional) form factors. For refinement, we pick the factor 2 (i.e. $M\to 2M$)  simply  for illustrative purposes. Extensions to any other factor appear to be possible, and we will assume that their fixed points are independent of the refinement choice (see the discussion in \cite{LLT3}). Similar to \cite{LLT1,LLT2}, from studying the discretised theories we deduce of how to define the discretised Hilbert spaces for the quantum theory and the {\it coarse graining maps} $J_{M\to M'}$  between Hilbert spaces of different resolution in the second subsection \ref{s32}. As the two introduced coarse graining maps -- called deleting kernel and filling kernel --
are fundamentally different, it is a priori not clear how the renormalisation group behaves with respect to both of them and whether both produce physically viable fixed points. To investigate this, we test both of them in section \ref{s33}, where we study a gauge-variant version of the toy model from the previous section -- hence not relying on a reduced phase space quantisation. This model features a Proca like mass term and higher powers of the Laplacian in order that holonomy operators be well-defined in the Fock space defined by that Hamiltonian. Hence, it gives first insights into theories allowing for holonomies and their renormalisation.
The fixed points can be found analytically after one adapts the coarse graining maps and chooses a suitable discretisation: While in the Fock representation induced by the continuum Hamiltonian holonomy operators do exist, as a first step we do not express the lattice approximants of the Hamiltonian in terms of lattice holonomies in order to simplify the analysis. In future work \cite{tbp}, in order to test the representation that is used in Loop Quantum Gravity, we aim at expressing the lattice Hamiltonians in terms of holonomies as well which makes the problem substantially more complicated as then the theory will be self-interacting.\\
In section \ref{s4_conclusion} we summarize our findings and conclude with outlook for further research.\\

\section{Reduced phase space quantisation for Abelian gauge theories}
\label{s2_ver0}
We present a possible strategy to extend the framework of direct Hamiltonian renormalisation developed in \cite{LLT1,LLT2,LLT3,LLT4} to Abelian gauge theories via reduced phase space quantisation. For this purpose, subsection \ref{s21} gives a short review of the framework as it was used for scalar fields.  The second subsection motivates a toy model in order to test the Hamiltonian renormalisation. To keep this preliminary study simple, we choose the Abelian gauge group ${\rm U}(1)$ and define the classical, continuum Hamiltonian in subsection \ref{s22} such that it resembles free Maxwell electrodynamics\footnote{
In the previous work \cite{BDH10} in addition to free scalar fields also free gauge theories such as Maxwell theory and linearised gravity were renormalised. While there are some similarities, the difference to the scalar field treatment of \cite{LLT2,LLT3,LLT4} and the present work is as follows: First, while \cite{BDH10} is concerned with the renormalisation of actions, we are concerned with renormalisation of vacua, Fock representations and Hamiltonians. Next, \cite{BDH10} provides explicit formulae for 1+1 dimensions while we treat 1+D dimensions for any D. Finally, \cite{BDH10} adapts the coarse graining map to the gauge symmetry while we perform a manifestly gauge invariant reduced phase space quantisation. With respect to the latter issue, see also \cite{tbp}.}. The actual computation of the renormalisation group flow is completely analogous to \cite{LLT2,LLT3,LLT4} and we will outline the general strategy in subsection \ref{s23}.

\subsection{Review: Classical discretisations of scalar fields}
\label{s21}
We consider an infinite dimensional, conservative Hamiltonian system defined on a globally hyperbolic spacetime of the form $\mathbb{R}\times \sigma$. If the spatial manifold $\sigma$ is not compact we introduce an infrared (IR) cut-off $R$ by restricting to smearing (i.e. test) functions which are defined on a compact submanifold, e.g. a torus $\sigma_R:=[0,R]^D$ if $\sigma=\mathbb{R}^D$. We will assume this cut-off $R$ to be implicit in all formulae below, but do not display it to keep them simple.\\
The dynamical variables of the system are the scalar field $\phi\in C^\infty(\sigma)$ and its canonical conjugated momentum $\pi_\phi$, i.e. $\{\pi_\phi(y), \phi(x)\}=\delta^D(x,y)$. We define their smearing against  test functions $f\in L(\sigma)$, i.e. functions from $\sigma$ to $\mathbb{R}$ whose properties we leave unspecified for the moment:
\begin{align}
\phi[f] := \int_\sigma {\rm d}^Dx\; \phi(x) \; f(x), \hspace{50pt}\pi_\phi[f]:=  \int_\sigma {\rm d}^Dx\; \pi_\phi (x) \; f(x)
\end{align}
Moreover, an ultraviolet cut-off $M$ is introduced in the form of some cell complex $C_M=\{c(m)\}_m$. The elements of the cell complex are regions $c(m)\subset \sigma$ such that $c(m)\cap c(m')=\emptyset$ and $\bigcup_m c(m)=\sigma$ and there are only finitely many elements, i.e. $|C_M|<\infty$. Knowledge of the $c(m)$ can be translated into knowledge of the indicator (or characteristic) functions $\chi_M(m):\sigma\to\{0,1\}$ which are defined as
\begin{align}
\chi_M(m)(x):=\begin{cases}
1 & {\rm if}\;\; x\in c(m)\in C_M\\
0 & {\rm else}
\end{cases}
\end{align}
Once a cell complex $C_M$ is chosen, one can introduce discretisations of the scalar field by restricting the observables (with respect to which the field is probed) to finite spatial resolution given by $C_M$ via the following choice of {\it evaluation} map:
\begin{align}\label{scalar_field_evaluation}
E_M\; :\hspace{5pt} L(\sigma)\;\; &\to\;\;  L_M\\
f\;\;&\mapsto\;\; f_M(m):=(E_M f)(m)=\epsilon_M^{-D}\int_\sigma {\rm d}^Dx\;f(x)\,\chi_M(m)(x)\nonumber
\end{align}
with $L_M$ being the set of finite sequences with $|C_M|$ many elements and $\epsilon_M^D=\int {\rm d}^Dx\, \chi_M(m)(x)$ which we assume to be independent of $m$ in the following. On the other hand, given a $f_M\,:\, \mathbb{Z}_{|C_M|}:=\{0,1,...,|C_M|-1\}\to \mathbb{R}$ we can embed it into the continuum via an {\it injection} map:
\begin{align}
I_M\;:\hspace{5pt} L_M\;\;&\to\;\; L(\sigma)\\
f_M\;\;&\mapsto\;\; (I_M f_M)(x):=\sum_m f_M(m)\chi_M(m)(x)=:f_M(\lfloor x\rfloor_{\chi_M})
\end{align}
We have introduced the map $\lfloor x\rfloor_{\chi_M} :=m$ such that $\chi_M(m)(x)=1$, which is always well-defined due to the properties of $C_M$. Defined in this way, $E_M$ serves as the left inverse of $I_M$:
\begin{align}
E_M \circ I_M ={\rm id}_{L_M}
\end{align}
Turning towards comparing discretisations of different resolutions with each other, we are mostly interested in families of cell complexes $\{C_M\}_M$ such that they define a partially ordered and directed set. This can happen, e.g., with defining $M< M'$ iff $\forall\; c'(m')\in C_{M'}$ there is $c(m)\in C_{M}$ such that $c'(m')\subset c(m)$.\footnote{By demanding that it is a proper subset, we guarantee that there are multiple elements in $C_{M'}$ forming a partition of $c(m)$.} It corresponds to viewing a function defined on coarse resolution as a function of finer resolution. Moreover, we restrict to finite partitions, meaning in particular that the number of cells $c'(m')$ contained in any $c(m)$ is finite: $N_{M',M}(m)<\infty$. (In \cite{LLT2,LLT3,LLT4} it was $N_{M',M}(m)=2^D$ for all $m$ but this simplification is merely to ease computations.) For the purpose of comparing different discretisations with each other, one introduces a map between the discretisations with respect to two cell complexes $C_M$ and $C_{M'}$ called {\it coarse graining} $I_{M\to M'}: L_M \to L_{M'}$ if $M<M'$. The coarse graining map is a free choice of the renormalisation group (RG) process whose flow it drives, and its viability can be tested only a posteriori. In \cite{LLT2,LLT3,LLT4} the main focus rested on choosing the concatenation of evaluation and injection for different discretisations as coarse graining:
\begin{align}\label{fillingkernel}
I_{M\to M'}=E_{M'}\circ I_M
\end{align}
However, let us mention that already in \cite{LLT3} also a second choice, called {\it deleting kernel}, was investigated: Let $M<M'$ and choose for any $m\in M$ a representative $m'_o(m)\in M'$ where $c'(m'_o(m))\subset c(m)$. Also, let $r$ be a mapping such that $r_{m,m'}=\delta_{m',m'_o(m)}$, i.e. selecting for $m'\in M'$ the representative $m_o'(m)$ of the coarse cell $c(m)$. Then
\begin{align}\label{deletingkernel}
(I^{Del}_{M\to M'}f_M)(m')= \sum_{m\in M} N_{M',M}(m) \;f_M(m) \; r_{m, m'}
\end{align}
In the quantum theory of free scalar fields both maps could be used to build injections that led to physically viable fixed point theories. However, it was only choice (\ref{fillingkernel}) which turned out to be {\it cylindrically consistent}, i.e.
\begin{align}
I_{M'}\circ I_{M\to M'} =I_M
\end{align}
Basically, this means that injection into the continuum can be done independently of the discretisation on which we consider the function to be defined, which is a physical plausible assumption.\\

We finish this section by presenting two examples for possible choices of cell complexes $C_M$ in case of the torus $\sigma_R=[0,R]^D$:\\
 {\bf (i) Discretisation using regular cubes.} The first example is the choice employed in \cite{LLT1,LLT2,LLT3,LLT4} which introduced a cubic lattice of $M\in\mathbb{N}$ points in each direction and with spacing $\epsilon_M=R/M$. Then, the characteristic functions of $C_M$ take the following form:
\begin{align}\label{cubiclattice}
^0\chi_M(m)(x) =\prod_{k=1}^{D} \bar{\chi}_M[m_k](x_k) \hspace{15pt}{\rm with }\hspace{15pt} \bar{\chi}_M[m_k](x_k)=\begin{cases}
1 & {\rm if}\;\; x_k\in [\epsilon_M m_k,\epsilon_M(m_k+1))\\
0 & {\rm else}
\end{cases}
\end{align}
However, this is by far not the only possibility. In order to demonstrate that nothing is special about the choice of tessellation of $\sigma$, we will use in section \ref{s3_ver1} the following cell complexes:\\
{\bf (ii) Discretisation using parallelepipeds.} 
We consider $D$-dimensional tessellations of the following form: at least one axis of the parallelepiped is aligned with one of the coordinate axes and a second axis of the parallelepiped connects diametral corners of an elementary hypercube. Then the remaining axes are either aligned with the coordinate axes or  explore all possibilities to connect diametral corners of lower dimensional hypercubes.
This yields 2 possibilities in $D$=2 and 9 possibilities in $D$=3. We can formalize this as follows: Let $i,j=1,...,D$ and
\begin{align}\label{parallelogramlattice}
&^{i,j}\chi_M(m)(x):=\int {\rm d}^Dy\; \delta(y_i-\epsilon_M m_i,\sum_{k\neq i} (y_k-\epsilon_M m_k))\Big(\prod_{k\neq i}\bar{\chi}_M[m_k](y_k)\Big)\times\\
&\hspace{150pt}\times \bar{\chi}_M[m_j](x_j-\sum_{k\neq j}(x_k-\epsilon_M m_k))\Big(\prod_{k\neq j} \delta(y_k,x_k)\Big)\nonumber
\end{align}
In $D$=2 the explicit form of the two possible parallelograms reads:
\begin{align}
^{1,1}\chi_M(m)(x)=\;^{2,1}\chi_M(m)(x)=\bar{\chi}_M[m_2](x_2)\;\bar{\chi}_M[m_1-m_2](x_1-x_2)\\
^{1,2}\chi_M(m)(x)= \;^{2,2}\chi_M(m)(x)=\bar{\chi}_M[m_1](x_1)\;\bar{\chi}_M[m_2-m_1](x_2-x_1)
\end{align}
In $D$=3 the fundamental cells take the form of parallelepipeds. While 9 different cases exist, we display only the explicit expressions for $i=1$:
\begin{align}
^{1,1}\chi_M(m)(x)&=\bar{\chi}_M[m_1-m_2-m_3](x_1-x_2-x_3)\;\bar{\chi}_M[m_2](x_2)\;\bar{\chi}_M[m_3](x_3)\\
^{1,2}\chi_M(m)(x)&=\bar{\chi}_M[m_2-m_1-m_3](x_2-x_1-x_3)\;\bar{\chi}_M[m_1-m_3](x_1-x_3)\;\bar{\chi}_M[m_3](x_3)\\
^{1,3}\chi_M(m)(x)&=\bar{\chi}_M[m_3-m_1-m_2](x_3-x_1-x_2)\;\bar{\chi}_M[m_1-m_2](x_1-x_2)\;\bar{\chi}_M[m_2](x_2)
\end{align}
and for $i=2,3$  similar functions with permutations of the indices are found.

\subsection{Phase space reduction of a continuum toy model}
\label{s22}
This subsection motivates and introduces a classical Hamiltonian system subject to the Gauss constraint for Abelian gauge group ${\rm  U}(1)$ in $D=3$ on a compact torus $\sigma=[0,1]^3$. The field content will be a ${\rm U}(1)$-connection $A_a$ and the corresponding electric vector field $E^a$. Due to ${\rm U}(1)$ being 1-dimensional, there is only one constraint per point, which reads:
\begin{align}\label{abelian-gauge-constraint}
G(x) = (\partial_a E^a)(x)
\end{align}
The most prominent example of a ${\rm U}(1)$ gauge theory is free Maxwell electrodynamics:
\begin{align}\label{Hamiltonian_electromagnetism}
H := \frac{\epsilon_o}{2}\int {\rm d}^3 x\; \big(E_a (x)E^a(x)+ c^2 A^\perp_a(x)\delta^{ab}(\omega^2 A^\perp)_b(x)
 \big)
\end{align}
with $A$ split into transversal and longitudinal part respectively:
\begin{align}\label{connection_perp}
A^{\perp}_a(x) = A_a(x)-A_a^{||}(x),\hspace{50pt} 
A^{||}_a(x)= (\partial_a\frac{1}{\Delta}\partial^b A_b)(x)
\end{align}
Further, $\epsilon_o$ is the electric constant of units $[J/(m V^2)]$, but in the following we set $c=\epsilon_o=1$. In Maxwell electrodynamics it is $\omega^2= -\Delta$ with $\Delta$ being the Laplacian. We modify (\ref{Hamiltonian_electromagnetism}) by replacing $\Delta$ with
\begin{align}\label{covariance}
\omega^2 = \frac{1}{p^{2(n-1)}}(-\Delta+p^2)^n
\end{align}
with some Proca like mass term $p>0$ and $n>0$. This is merely a generalisation as standard Maxwell theory can be reobtained in the limit $p\to 0$ and $n=1$.\\

Our goal is to go to the reduced phase space and therefore we also split the electric field $E^a$ into $E^a_\perp$ and $E^a_{||}$ defined similar to (\ref{connection_perp}):
\begin{align}\label{electric_perp}
 E_{\perp}^a(x) = E^a(x)-E^a_{||}(x),\hspace{50pt}
E_{||}^a(x)= (\partial^a\frac{1}{\Delta}\partial_b E^b)(x)
\end{align}
Due to the fact that the transverse modes are gauge-invariant, i.e. $ \{G[\Lambda], A^{\perp}_a(x)\}$ $=\{G[\Lambda], E^a(x)\}=0$ for all $\Lambda\in L(\sigma)$, it follows that the Hamiltonian (\ref{Hamiltonian_electromagnetism}) is gauge-invariant, too.\\
The unreduced phase space is equipped with Poisson brackets $\{E^a(x),A_b(y)\}=\delta^a_b \delta^{(3)}(x,y)$. As is standard, we perform a canonical transformation to:
\begin{align}
\{E^a_{||}(x),A^{||}_b(y)\}=\frac{1}{\Delta}\partial^a&\partial_b \delta^{(3)}(x,y),\hspace{30pt}
\{E^a_{\perp}(x),A^{\perp}_b(y)\}=(\delta^a_b-\frac{\partial^a\partial_b}{\Delta})\delta^{(3)}(x,y)\nonumber
\\
&\{E_{\perp}^a(x),A^{||}_b(y)\} = \{ E_{||}^a(x), A^{\perp}_b(y)\}=0
\end{align}
Next, we reduce to the subspace $A^{||}=E_{||}=0$ and go into Fourier space
\begin{align}
\hat{E}^a_{\perp}(k):=\int {\rm d}^3x\; e^{i\, k\cdot x} E^a_{\perp}(x),\hspace{40pt} \hat{A}^{\perp}_a(k):=\int {\rm d}^3 x\; e^{i\,k\cdot x} A^{\perp}_a(x)
\end{align}
which can be decomposed as
\begin{align}
\hat{E}^a_{\perp}(k) = \epsilon^a_1 (k) \hat{\rm E}_1(k) + \epsilon^a_2(k)\hat{\rm E}_2(k), \hspace{40pt} \hat{A}^a_{\perp}(k) = \epsilon^a_1 (k) \hat{\rm A}_1 (k) + \epsilon^a_2(k) \hat{\rm A}_2(k)
\end{align}
with a choice of vector fields $\epsilon_1(k),\epsilon_2(k)$ which are orthonormal to each other and orthogonal to $k^a$. Such a choice can always be made and implies that the symplectic structure between $\hat{\rm E},\hat{\rm A}$ is of canonical form, i.e. for $I,J\in \{1,2\}$ 
\begin{align} \label{final_phase_space}
\{ \hat{\rm E}_I(k), \hat{\rm A}_J (k') \} = \delta_{IJ} \delta^{(3)}(k,k')\,.
\end{align}
On this subspace the Gauss constraint is trivially solved, and all gauge-degrees of freedom have been removed. Expressed in these variables the continuum Hamiltonian of our model takes the form:
\begin{align}\label{final_test_Hamiltonian}
H=\int {\rm d}^3 k\; \sum_{I=1,2} \big(
|\hat{\rm E}_I(k)|^2+\omega(k)^2 |\hat{\rm A}_I(k)|^2
\big)
\end{align}

\subsection{Scalar field renormalisation with multiple field species}
\label{s23}
In this subsection we discretise the model (\ref{final_test_Hamiltonian}) with $\omega$ from (\ref{covariance}) with the scalar field techniques introduced in \cite{LLT1}. Due to the form of the Hamiltonian we are close to the analysis in \cite{LLT2,LLT3,LLT4} to which we refer the reader for all details. Indeed, we can understand the Hamiltonian as two decoupled field species $({\rm E}_I, {\rm A}_I)$ labelled by $I=1,2$, where we use the Fourier inversion:
\begin{align}
{\rm E}_I(x):= \frac{1}{2\pi} \int {\rm d}^3k \; e^{-i \;k\cdot x} {\rm \hat{E}}_I(k)  ,\hspace{40pt} {\rm A}_I (x) :=\frac{1}{2\pi} \int {\rm d}^3k \; e^{-i \;k\cdot x} {\rm \hat{A}}_I(k)  
\end{align}
We introduce a family of discretisations of the spatial manifold $\sigma$ in terms of cubic cell complexes as described in the previous subsection such that $N_{M,2M}(m)=2^D$ for all $m\in\mathbb{Z}_M^3=\{0,1,...,M-1\}^3$. With the evaluation maps $E_M$ from (\ref{scalar_field_evaluation}) we discretise both field species:
\begin{align}
{\rm E}_{M,I}(m):= {\rm E}_I(m \epsilon_M),\hspace{50pt} {\rm A}_{M,I}(m):= \epsilon_M^3 (E_M {\rm A}_I)(m)
\end{align}
We must also introduce a discretisation of $\omega$ which is supposed to map from $L_M\to L_M$. Since we have two field species $I=1,2$ it could turn out that each supports its own covariance. To take this possibility into account, we will keep the discretisations $\omega_{M,I}$ dependent on the field species $I$ in the following. However, as initial discretisation we take them to be equal, that is:
\begin{align}
\omega_{M,1} = \omega_{M,2} = \omega_M^{(0)} \equiv \frac{1}{p^{2(n-1)}}(-\Delta_M +p^2)^n
\end{align}
with $\Delta_M$ some initial discretisation such that $\lim_{M\to \infty} \omega_M^{(0)}=\omega$.\\
Since the Hamiltonian is essentially of free harmonic oscillator form for each $I$, it motivates to introduce the discrete annihilation and creation fields:
\begin{align}\label{disc_annihilation1}
{\bf a}_{M,I}(m) := \frac{1}{\sqrt{2\hbar}}\big(\sqrt{\omega_{M,I}/ \epsilon_M^3} {\rm A}_{M,I}(m) -  i\sqrt{ \epsilon_M^3 / \omega_{M,I} } {\rm E}_{M,I}(m)]\big)
\end{align}
such that
\begin{align}\label{disc_Hamiltonian1}
H_M = \hbar \sum_{I=1,2} \sum_{m\in\mathbb{Z}^3_M}  \overline{{\bf a}_{M,I}}(m)(\omega_{M,I} {\bf a}_{M,I})(m)
\end{align}
For any resolution $M$ we define the corresponding Hilbert spaces $\mathcal{H}_{M,I}$ for specie $I$ with Fock vacuum $\Omega_M^{(0)}\in\mathcal{H}_M:=\otimes_I \mathcal{H}_{M,I}$ annihilated by the operators corresponding to (\ref{disc_annihilation1}), i.e.
\begin{align}
\widehat{{\bf a}_M}[f_M]:=\sum_I \sum_{m\in\mathbb{Z}^3_M} \hat{\bf a}_{M,I}(m) f_{M,I}(m)
\end{align}
(with $f_{M,I}(m)\in L_M^2:=\ell_2(\mathbb{Z}_M^{3})^2$). Thus, $\Omega_M$ is simultaneously annihilated by the quantisation of (\ref{disc_Hamiltonian1}).
Denoting by $\langle .,.\rangle_{\mathcal {H}_M}$ the scalar product on $\mathcal{H}_M$ it follows 
\begin{align}
\langle \Omega_M, e^{i \widehat{ {\rm A}_{M}}[f_M]} \Omega_M\rangle_{\mathcal{H}_M} = e^{-\frac{\hbar}{4}\langle f_{M,1},\omega_{M,1}^{-1} f_{M,1}  \rangle}\; e^{-\frac{\hbar}{4}\langle f_{M,2},\omega_{M,2}^{-1} f_{M,2} \rangle}
\end{align}
Each $\mathcal{H}_{M}$ can be represented as Hilbert space $L_2(\mathbb{R}^{2M^3},d\nu_M)$ where $\nu_M=\nu_{M,1}\nu_{M,2}$ is a Gaussian measure with covariance $c_M={\rm diag}(c_{M,1}, c_{M,2})$ and $c_{M,I}=\frac{\hbar}{2}\omega_{M,I}^{-1}$. Hence, we have at our disposal an initial family of Osterwalder-Schrader data $(\mathcal{H}_M, \hat{H}_M,\nu_M)$ which under a renormalisation step, does not change its general structure \cite{LLT2} but leads to a new family of (Gaussian) covariances, i.e. $\{c_{M,I}^{(n)}\}_{M} \to \{c_{M,I}^{(n+1)}\}_M$. Our goal is to find  a family of measures that remains invariant under the coarse graining induced by the maps $I_{M\to 2M}$ defined in subsection \ref{s21}.\\

Indeed, the fact that our model is essentially two copies of a free scalar field allows making use of many tools developed in \cite {LLT2,LLT3,LLT4}. We recall from section 3.1 of \cite{LLT4} that determination of the fixed points for any power $n$ in (\ref{covariance}) can be reduced to studying the renormalisation group flow for $n=2$ at the cost of an additional contour integral by a standard application of the residue theorem: Starting from the initial covariance:
\begin{align}\label{inverseresidualtheorem}
c_M^{(0)} = \frac{ p^{(n-1)}\hbar/2}{(-\Delta_M+p^2)^{n/2}}=\frac{p^{(n-1)}\hbar/2}{2\pi i}\oint_{\gamma}{\rm d}k_o\;\frac{1}{k_o^{n/2}}\;\;\frac{1}{-\Delta_M+p^2-k_o}
\end{align}
with $\gamma$ being a contour consisting of a part along $i\mathbb{R}$ (excluding the origin) and  an arc closing at infinity on the positive half plane. For brevity, we relabel $\bar{p}^2:=p^2-k_o$. Now, since the RG flow is linear and only changes $\Delta_M$, determination of the fixed points boils down to the case $n=2$ up to said contour integral along $\gamma$.\\
As we had already seen in \cite{LLT2} that the RG flow is easiest studied in the Fourier transformed representation, we recall the discrete Fourier transform and its inverse on $L_M$ for any $D$ (with $k_M=\frac{2\pi}{M}$)
\begin{align}\label{disc_FT}
f_M(m)=\sum_{l\in\mathbb{Z}^D_M} \hat{f}_M(l) e^{i k_M l\cdot m},\hspace{40pt}
\hat{f}_M(l) = M^{-D} \sum_{m\in\mathbb{Z}^D_M} f_M(m) e^{-ik_Ml\cdot m}
\end{align}
Going to the discrete Fourier picture and assuming translational invariance of the covariance, we know that the kernel of the covariance at the fixed point can be written as:
\begin{align}
c_{M,I}^*(m) = \frac{p^{(n-1)}\hbar/2}{2\pi i}\oint_{\gamma}{\rm d}k_o\;\frac{1}{k_o^{n/2}}\; \sum_{l\in\mathbb{Z}^3_M}e^{i\; k_Ml\cdot m}\; \hat{c}^*_{M,I}(l)
\end{align}
Further, it was observed in \cite{LLT4} that the  renormalisation group flow decouples for each direction and thus the covariance can be transformed via another application of the residue theorem into:
\begin{align}\label{facotrisation_s2}
\hat{c}_{M,I}^{*}(l) = \frac{1}{(2\pi i)^3}(\prod_{b=1}^3\oint_\gamma {\rm d}z_b) \; \frac{1}{\bar{p}^2-\sum_b z_b} \prod_{b=1}^3  \hat{c}^*_{M,I,b}(l_{b};z_b)
\end{align}
For $I_{M\to 2M}$ from (\ref{fillingkernel}) with a discretisation using regular cubes as (\ref{cubiclattice}) the fixed point obtained from the flow starting with the fraction in (\ref{inverseresidualtheorem}) has been already computed in \cite{LLT2} and reads:
\begin{align}\label{standard_blocking_fp}
\hat{c}^*_{M,I,b}(l,z)=\hat{\underline{c}}_{M}(l,q):= \frac{\epsilon_M^2}{q^3}\frac{q\, {\cosh}(q)-{\sinh}(q)+({\sinh}(q)-q)\cos(k_M l)}{{\cosh}(q)-\cos(k_M l)}
\end{align}
where $q=\epsilon_M \sqrt{z}$. Note that indeed $I_{M\to2M}$ is the same in each direction $b$ and the same for both field species $I$, hence we obtain the same fixed point for both $I$.\\
For the deleting kernel $I^{Del}_{M\to 2M}$ from (\ref{deletingkernel}) the fixed point can be computed to be\footnote{Note that the earlier
work \cite{LLT3} contains a typo: While in eqn (3.61) (in \cite{LLT3}) we quote obviously the initial covariance, we missed to explicitly write the fixed point given by (\ref{standard_del_fp}) above.}
\begin{align}\label{standard_del_fp}
\hat{c}^*_{M,I,b}(l,z)=\hat{c}^{Del}_{M}(l,q):=\frac{\epsilon_M^2}{q}\frac{{\rm sinh}(q)}{{\rm cosh}(q)-\cos(k_M l)}
\end{align}
Thus, we finished the analysis of the direct Hamiltonian Renormalisation applied to our toy model for a gauge theory which has been reduced to the gauge-invariant subspace before quantisation. Keep in mind that in section  3.2.2 of \cite{LLT2} it was already explained that renormalisation of the Hamiltonian leads to replacing in the discretisation (\ref{disc_Hamiltonian1}) the initial covariance with the fixed pointed one, that is $\omega_{M,I} \mapsto \omega^*_M$.\\
Also, since both field species behaved exactly the same, i.e. $\omega_{M,I}=\omega_M$, the same universality and continuum properties discussed in \cite{LLT3,LLT4} apply to this case as well.

\section{Renormalisation with form factors for free vector bosons}
\label{s3_ver1}

In this section we turn towards those discretisations for which the fields are discretised with respect to the edges of some finite graph. This brings us closer to lattice gauge theories which are typically formulated in terms of holonomies, that is exponentials of the connection. For this purpose, subsection \ref{s31} introduces discretisations where the fields are integrated along one-dimensional curves and their canonical conjugated pairs against $D-1$ faces, where $D$ is the number of spatial dimensions. We can express the discretisation in a language maximally close to \cite{LLT3} and the previous section, if we smear both objects with {\it form factors} of curves and $D-1$ faces respectively.\\
Due to our earlier considerations we have an understanding how sensible injection maps on the quantum level can be chosen, which we do in subsection \ref{s32} calling them ``deleting'' and ``filling'' kernel respectively. These relate the quantities of some resolution $M$ to those on a finer resolution $pM$, where $p\in\mathbb{N}$ can be any arbitrary factor. However, to keep the notation simple, we will use throughout this paper the choice $p=2$.\\
Afterwards, we want to investigate a toy model in order to test how the different coarse graining maps and their corresponding fixed pointed theories behave with respect to each other. As we want to study models which allow for the existence of holonomy operators in the Fock representation that supports the continuum Hamiltonian, we have to introduce a deformation of free Maxwell theory. This deformation is discussed in subsection \ref{s33}.\\
In subsection \ref{s34_del} and \ref{s35_fil} we will again employ tools developed in \cite{LLT2,LLT3,LLT4} to determine the fixed pointed Hilbert spaces for the coarse graining maps defined by the deleting as well as the filling kernel. The task amounts to finding a suitable fixed pointed covariance defining a Gaussian measure on the Hilbert spaces of finite resolution, which we will derive in closed form for both maps. This demonstrates robustness of the continuum theory even under drastic changes of the coarse graining procedure.

\subsection{Injection and evaluation maps}
\label{s31}

As in the previous section, we consider a ($D$+1)-dimensional manifold of the form $\mathbb{R}\times \sigma$ on which an infinite dimensional, conservative Hamiltonian system is defined. Via an IR cut-off we restrict to the compact submanifold $\sigma_R$, omitting the cut-off $R$ in all subsequent formulas.\\
Let the phase space be coordinatised by vector fields $E^a$ and covector fields $A_a$ with $a=1,...,D$ which read in terms of smearing against test functions $f,g\in L(\sigma)^D$ 
\begin{align}
E[g]&:=\langle E,g\rangle:= \int_\sigma {\rm d}^D x\; E^a(x) g_a(x)\\
A[f]&:=\langle A,f\rangle:=\int_\sigma {\rm d}^D x \; A_a(x) f^a(x) \label{integratedA}
\end{align}
and which have elementary Poisson brackets: 
\begin{align}\label{sympl_Struc}
\{E[g],E[g']\}=\{A[f],A[f']\}=0,\hspace{30pt}\{E[g],A[f]\}=\kappa_o \langle g,f\rangle :=\kappa_o \int_\sigma {\rm d}^Dx\; g_a(x) f^a(x)
\end{align}
with $\kappa_o$ being the coupling constant of the theory, which we set to one in the following: $\kappa_o=1$.\\
We discretise the theory by introducing smearings of $A_a$ along the 1-dimensional edges of some dual cell complex. For the case of $\sigma_R=[0,R]^D$, it suggests itself to consider regular lattices, where at each vertex there are 2D many edges incident. In the following, we will restrict to this choice, to keep the notation simple. Note that the edges of the lattice are understood to be {\it paths}, i.e. semianalytic curves. The set of all paths forms the groupoid $\mathcal{P}$, which is closed under concatenation of elements and features an inverse element for each path -- however there is no natural identity element on $\mathcal P$. We understand an element $e\in\mathcal{P}$ as the embedding $e:[0,1]\to \sigma_R$. Since we want to focus for the purpose of this article on regular lattices (e.g. cubic lattices for $D=3$), we are mostly interested in a subset of $\mathcal{P}$: Given a lattice $\gamma_M$, where $M$ denotes the number of vertices in each direction, we denote the set of oriented edges in $\gamma_M$ by $\mathcal{P}^M\subset \mathcal{P}$.

A smearing of the field $A_a$ against an edge can be obtained by allowing in (\ref{integratedA}) not only test functions in $L(\sigma)$ but distributions such as {\it form factors} $F_e$ for any edge $e\in \mathcal P^M$, i.e. :
\begin{align}\label{formfactor_def}
(F_e)^a(x)=\int_e {\rm d}y^a \delta^{(D)}(x,y)
\end{align}
Similarly, since we are interested in those lattices $\gamma_M$ which stemmed from some dual cell complex, we can associate with each edge $e$ a choice of some $(D-1)$-dimensional face $S(e)$, such that $S(e)\cap e' = \emptyset$ iff $e \neq e'$ and at the unique point $S(e)\cap e$ its normal points in the same direction as $\dot{e}$. Then, we can also introduce the dual form factors of the face $S$, e.g.:
\begin{align}
({\rm f}^S)_a(x)=\int_S{\rm d}y^b\wedge {\rm d}y^c\;\frac{\epsilon_{abc}}{2}\;\delta^{(3)}(x,y)\hspace{30pt}{\rm if}\;D=3\\
({\rm f}^S)_a(x)=\int_{S}{\rm d}y^b\;\epsilon_{ab}\; \delta^{(2)}(x,y)\hspace{82pt}{\rm if}\;D=2
\end{align}
Note that there is a natural non-distributional Poisson bracket between the form factors for curves and the dual form factors for faces:
\begin{align}\label{formfactor_delta}
\langle F_e, {\rm f}^{S(e')}\rangle= \delta_{ee'}
\end{align}
We can now restrict the set of our observables with respect to which the physical configuration $(E^a,A_a)$ is probed. We want to keep only those observables that can be understood as restricting $A_a$ to the edges of a lattice and $E^a$ to its dual faces. This can be achieved by introducing injection and evaluation maps between test functions in $L(\sigma^D)$ and functions on the lattice $L_M\equiv L(\mathcal P^M, \mathbb{R})$:
\begin{align}
\label{deleting_test_functions_injection}
I_M^{Del} \; : \;\; L_M &\hspace{10pt}\to \hspace{10pt} L(\sigma)^D\\
 f_M &\hspace{10pt}\mapsto \hspace{10pt} (I_M^{Del} f_M)^a (x):=  \sum_{e\in\mathcal P^M} (F_e)^a(x) f_M(e)\nonumber\\
\label{deleting_test_functions_evaluation}
 E_M' \,: \, L(\sigma)^D &\hspace{10pt}\to \hspace{10pt} L_M\\
 f^a &\hspace{10pt}\mapsto \hspace{10pt} (E_M'f)(e):= \langle f,{\rm f}^{S(e)}\rangle \nonumber
\end{align}
Using property (\ref{formfactor_delta}) one easily verifies that $E_M'\circ I_M^{Del} ={\rm id}_M$. Further, we can understand
\begin{align}\label{smearings_to_discconf}
A[I_M^{Del} f_M] =\int{\rm d}^3x\; A_a(x) (I_M^{Del} f_M)^a(x)=\sum_{e\in\mathcal P^M} A[F_e]\; f_M(e)
\end{align}
as $A_a$ restricted to the lattice $\gamma_M$. We introduced a superscript on $I^{Del}_M$ and call it in the following ``deleting kernel'' due to its similarity with (\ref{deletingkernel}).\footnote{Deleting kernels are favoured in the literature on cylindrical consistency of gauge theories, see for example the projective spaces of the Ashtekar-Lewandowski Hilbert space in the context of Loop Quantum Gravity \cite{Thi07,AL95,Rov04}. Note however, that the Ashtekar-Lewandowski Hilbert space for each edge is a Hilbert space over ${\rm SU}(2)$ in contrast to the Fock space we consider in this manuscript.}
Yet, this construction is far from unique and in order to demonstrate this we introduce a second choice called ``filling kernel''. In its spirit, this map is constructed to be similar to the standard choice employed for scalar fields, i.e. (\ref{scalar_field_evaluation}). Due to the multiple choices of cell complexes used to define (\ref{scalar_field_evaluation}), we have an ambiguity regarding the injection map for the ``filling kernel''. 
We restrict us to the choices of parallelepipeds (\ref{parallelogramlattice}) since discretisations with regular cubes have been extensively studied in the papers \cite{LLT2,LLT3,LLT4} and this new choice will demonstrate the robustness of the renormalisation procedure under considerably drastic changes. For ${\rm i}\in\{1,...,D\}$, we define
\begin{align}
I_M^{\rm Fil,i} \; : \;\; L_M &\hspace{10pt}\to \hspace{10pt} L(\sigma)^D\\
 f_M &\hspace{10pt}\mapsto \hspace{10pt} (I_M^{\rm Fil,i}f_M)^a(x)=\frac{1}{2^{D-1}}\sum_{e\in\mathcal P^M}\sum_{e'\in\mathcal P^{2M}} (F_{e'})^a(x)f_M(e) r_{ee'}^{\rm Fil,i}
\end{align}
with 
\begin{align}\label{representative_mappings1}
 r^{\rm Fil,i}_{ee'}=\begin{cases}
\delta_{\dot{e}(0),\dot{e}'(0)}\delta_{\dot{e}(0),\hat{b}}& {\rm if}\; {\rm Im}(e')\subset \;^{{\rm i},b}\chi(e(0))\\
0 & {\rm else}
\end{cases}
\end{align}
where ${\rm Im}(e')$ denotes the image of $e':[0,1]\to \sigma$, $^{{\rm i},b}\chi$ is defined in (\ref{parallelogramlattice}), $\hat{b}$ is the normal vector of unit length pointing in direction $b$ and $\delta_{\dot{e}(0),\hat{b}}$ denoting the Kronecker delta in the tangent space, i.e. it is non-vanishing only if $\dot{e}(0)=\hat{b}$. Note that the cases in (\ref{representative_mappings1}) are meant to be checked for all possible $b=1,...,D$ separately.\\
It is easy to check that for a suitable choice of faces $S(e)$ we get $E'_M\circ I^{\rm Fil, i}_M={\rm id}_M$ for all $\rm i$. We recall that the parameter $\rm i$ of the filling kernel determines the choice of parallelepipeds from (\ref{parallelogramlattice}) and thus all derived quantities in the coarse graining procedure will depend on it. In what follows we fix ${\rm i}\in \{1,2,3\}$ and check the coarse graining maps for all of them separately, thus not displaying the label ${\rm i}$ explicitly.

\subsection{Coarse graining for deleting and filling kernel}
\label{s32}
In this subsection we concatenate injection and evaluation maps to {\it coarse graining maps} $I_{M\to 2M}$ both for deleting and filling kernel on the classical level and use $I_{M\to 2M}$ to build isometries between Fock quantised Hilbert spaces of different resolutions.

\subsubsection{Classical coarse graining maps}
First, we introduce the coarse graining maps for the deleting kernel from $L_M\to L_{2^n M}$ via:  ($n\in\mathbb{N}$)
\begin{align}\label{CG_Del}
I_{M\to 2^nM}^{Del}:=(\frac{\epsilon_{2^n M}}{\epsilon_M})^{\alpha} E_{2^n M}\circ 
I^{Del}_M
\end{align}
with $\alpha \in \mathbb{R}$. They relate a set of test functions on coarse resolution $M$ with a set of test functions at finer resolution $2^n M$. Their action on test functions can be written explicitly as:
\begin{align}\label{expclit_Del}
f_{2^n M} (e') := (I^{Del}_{M\to 2^nM}\; f_M)(e') =\frac{1}{2^{n\,\alpha}} \sum_{e\in\mathcal{P}^M}r_{ee'}^{Del}\;  f_M(e)
\end{align}
where $e'\in \mathcal{P}^{2^n M}$ and
\begin{align}\label{representative_mappings2}
r^{Del}_{ee'}=\begin{cases}
1 & {\rm if} \; e'\subset e\\
0 & {\rm else}
\end{cases}.
\end{align}
The free parameter $\alpha\in\mathbb{R}$ can be chosen in such a way that the condition of {\it cylindrical consistency} is satisfied, that is for all $A$ and $f_M\in L_M$: 
\begin{align}
A[I_{2^n M}\circ I_{M\to2^n M} f_M ]=A[I_M f_M]
\end{align}
Using that $F_e=F_{e_1}+F_{e_2}+...+F_{e_{2^n}}$ if $e=e_1\circ e_2\circ ...\circ e_{2^n}$ we find
\begin{align}\label{comp_s32}
A[I_{2^n M}^{Del}I_{M\to2^n M}^{Del} f_M]&=\int {\rm d}^Dx \sum_{e'\in \mathcal P^{2^nM}} A_a(x)(F_{e'})^a(x)\left(2^{-\alpha n}\sum_{e\in\mathcal{P}^M} f_M(e) r_{ee'}^{Del}\right)=\nonumber\\
&=\int {\rm d}^Dx \sum_{e\in \mathcal{P}^M} A_a(x) f_M(e) \left(
2^{-\alpha n}\sum_{e'\in \mathcal{P}^{2^n M}}(F_{e'})^a(x) r^{Del}_{ee'}\right)=\nonumber\\
&=\int {\rm d}^Dx \sum_{e\in \mathcal{P}^M} A_a(x)2^{-\alpha n} (F_e)^a(x) f_M(e) =A[I_M^{Del} f_M] 2 ^{-\alpha n}
\end{align}
Hence, it must be $\alpha=0$.\\

If we were to introduce a coarse graining map of the filling kernel as the analogue of (\ref{CG_Del}), a calculation similar to (\ref{comp_s32}) demonstrates, that the latter is {\it not } cylindrical consistent unless $A_a$ is constant over each $^{{\rm i},b}\chi$.
However, requiring cylindrical consistency for the classical coarse graining map is not necessary per se, thus this finding does not rule out the filling kernel. The important property for the inductive limit construction is the compatibility condition between the quantum isometries, which follows from the weaker condition
\begin{align}\label{weaker_cylconsistency}
I_{2^n M\to 2^{n'} M} I^{\textcolor{white}{A^n_n}}_{M\to 2^n M} = I_{M \to 2^{n'} M} 
\end{align}
with $n<n'\in\mathbb{N}$. Indeed, (\ref{weaker_cylconsistency}) can be achieved also for the filling kernel when defining $I_{M\to2^nM}$ as the analogue of (\ref{expclit_Del}):
\begin{align}
f_{2^n M} (e') := (I^{\rm Fil,i}_{M\to 2^nM}\; f_M)(e') =\frac{1}{2^{n}} \sum_{e\in\mathcal{P}^M}r^{\rm Fil,i} _{ee'}\;  f_M(e)
\end{align}
with $e'\in \mathcal{P}^{2^nM}$ and $r^{\rm Fil,i}_{ee'}$ from (\ref{representative_mappings1}).

Lastly, it turns out -- for both filling and deleting kernel -- that demanding  the map $I_{M\to2^n M}$ to be an isometry, i.e.
\begin{align}
\langle I_{M\to2^n M} f_M,I_{M\to2^n M}\tilde{f}_M\rangle_{2M}=\langle f_M,\tilde{f}_M\rangle_M,\hspace{30pt} \forall f_M,\tilde{f}_M
\end{align}
can be used to fix an auxiliary scalar product on $L_{M}$:
\begin{align}\label{scalarproduct_testfunctions}
\langle f_M, \tilde{f}_M\rangle_M := \epsilon^D_M \sum_{e\in\mathcal{P}^M} f_M(e) \tilde{f}_M(e)
\end{align}

\subsubsection{Isometric injections on the quantum level}
In this section we construct coarse graining maps between Hilbert spaces corresponding to different resolutions. These maps drive the renormalisation group (RG) flow between the inner products on the Hilbert spaces $\mathcal{H}_M$. Once a fixed point family of Hilbert space measures is found, it can be used to obtain a continuum Hilbert space via the method of inductive limits \cite{Jan88,KR86}. To use the latter toolbox, certain requirements must be met for the coarse graining maps $J_{M\to2M}$: It must be guaranteed that $J_{M\to 2M}$ are isometric injections, i.e.
\begin{align}
J_{M\to 2M}^\dagger J_{M\to 2M}^{\textcolor{white}{2_{n_n}}} = {\rm id}_{\mathcal{H}_M}
\end{align}
and that they are subject to the compatibility condition, i.e. for each $n<n'\in\mathbb{N}$:
\begin{align}\label{compatibility_condition}
J_{2^nM\to2^{n'}M} \circ J_{M\to 2^nM}^{\textcolor{white}{2_{n_n}}} = J_{M\to 2^{n'} M}
\end{align}
These two properties were also imposed for scalar field models and indeed the same procedure of constructing the injections from \cite{LLT1} can be used again. We utilize a Fock quantization of the discretised field $A_M=A\circ I_M$. Upon choosing the vacuum vector $\Omega_M\in\mathcal{H}_M$ of the discretized Hamiltonian, we consider the dense linear span of vectors of the form
\begin{align}\label{generating_functional_Abelian}
\exp(i\hat 	A_M [f_M]) \;\Omega_M 
\end{align}
where
\begin{align}
\hat{A}_M[f_M] := \epsilon_M \sum_{e\in\mathcal{P}^M} \hat{A}_M (e) f_M(e) =\epsilon_M \sum_{m\in\mathbb{Z}_M^D}\sum_{a=1}^D \hat{A}_{M,a}(m) f_{m,a}(m)
\end{align}
and we denote the edge $e=e_{m,a}$ with initial vertex $m$ and direction $a$.\\
In the same manner as in \cite{LLT1}, we define the injections between Fock spaces as:
\begin{align}\label{quantum_coarse_graining}
J_{M\to 2M} (e^{i\hat A_M[f_M]}\Omega_M) := e^{i\hat A_{2M} [I_{M\to2M} f_M]}\Omega_{2M}
\end{align}
where $I_{M\to2M}$ is the respective version of its action on test functions for deleting or filling kernel.\\
By construction, this map is maximally parallel to the case of scalar fields and therefore many properties can be transferred to this setting. We refer to \cite{LLT1,LLT2} for further details.\\

\subsection{Toy model: Definition and discretisation for a Proca like theory}
\label{s33}

In this subsection, we define a toy model which allows for holonomy like operators in the continuum, i.e. $\exp(i \hat{A}[F_\alpha])$ has finite expectation values for $\alpha$ being some closed curve in $\sigma$. Then, we discretise this theory with respect to smearings along the curves of a lattice $\gamma_M$ as discussed before.

\subsubsection{Definition of the continuum model}

In close analogy to the model of section \ref{s23} we study a field theory with $D$=3 spatial dimensions and Hamiltonian
\begin{align}\label{Hamiltonian_vectorbosons}
H := \frac{\epsilon_o}{2}\int {\rm d}^3 x\; \big(E_a (x)E^a(x)+ c^2 A^a(x)(\omega^2 A)_a(x) \big)
\end{align}
where in the following we set $c=\epsilon_o=1$. In order to allow for the continuum QFT to support the exponentials of Wilson loops as operators, i.e. $\exp(i \hat{A}[F_\alpha])$ with some closed curve, $\alpha$  we chose 
\begin{align}\label{covariance_s33}
\omega^2 = \frac{1}{p^{2(n-1)}}(-\Delta+p^2)^n
\end{align}
with some mass term $p>0$ and $n\geq4$ to ensure existence of the covariance following from (\ref{Hamiltonian_vectorbosons}) when evaluated on form factors $F_\alpha$ as in (\ref{formfactor_def}): \\

\noindent{\bf Lemma:} Let $\alpha:[0,1]\to\sigma$ be a (closed) curve. The continuum vacuum expectation value of the holonomy along $\alpha$ is finite if $n\geq 4$ and $p>0$, i.e.:
\begin{align}
\langle \Omega, e^{i \hat{A}[F_{\alpha}]} \Omega \rangle_{\mathcal{H}}=e^{-\hbar\langle F_\alpha,\, \omega^{-1} F_\alpha \rangle/4} < \infty
\end{align}\\
{\bf Proof:} We consider only the case $n=4$ as higher powers are automatically included due to positive definiteness of $-\Delta$ and $p>0$. The vacuum expectation value will be finite if $\langle F_\alpha, \omega^{-1} F_\alpha \rangle$ remains finite with $\omega$ from (\ref{covariance_s33}). It suffices to check whether
\begin{align}
\int  \frac{{\rm d}^3k}{(p^2+k^2)^2} \; (\hat{F}_\alpha)^a(k)\overline{(\hat{F}_\alpha)_a(k)}  < \infty
\end{align}
where
\begin{align}
(\hat{F}_\alpha)^a(k)&=\int {\rm d}^3x\; e^{i\,k\cdot x} (\hat{F}_\alpha)^a (x)=\int_{\alpha } {\rm d}y^a \int {\rm d}^3x \; e^{i\,k\cdot x} \delta^{(3)}(x,y)\nonumber\\
&=\int_\alpha {\rm d}y^a e^{i\,k\cdot y}= \int_0^1 {\rm d}t\; \dot{\alpha}^a(t) e^{i\,k\cdot \alpha(t)}
\end{align}
First, we give a bound from above for the absolute value of $\hat{F}_\alpha$
\begin{align}
|| \hat{F}_\alpha(k) || \leq \beta := \sup_{a=1,2,3;t\in [0,1]} |\dot{\alpha}^a(t)|
\end{align}
Using this approximation and going to spherical coordinates ${\rm d}^3k \to r^2 \sin(\theta)\,{\rm d}r \,{\rm d}\theta\, {\rm d}\varphi$ we get:
\begin{align}
\int_{\mathbb{R}^3}  \frac{{\rm d}^3k}{(p^2+k^2)^2} &\; (\hat{F}_\alpha)^a(k)\overline{(\hat{F}_\alpha)_a(k)}\leq  \beta^2 \int_{\mathbb{R}^3 } \frac{{\rm d}^3k}{(k^2+p^2)^2}=4\pi \beta^2 \int_0^\infty  \frac{r^2\;{\rm d}r}{(r^2+p^2)^2}=\nonumber\\
&=2\pi \beta^2 \int_{\mathbb{R}} \frac{r^2\;{\rm d}r }{(r^2+p^2)^2}
=\frac{2\pi \beta^2}{p}  \int_{\mathbb{R}} \frac{ x^2\;{\rm d}x}{(x^2+1)^2} =\frac{\pi^2\beta^2}{p} < \infty\label{proof_1}
\end{align}
where we used the residue theorem in the last step. Hence, the vacuum expectation value is well-defined. \qed
 \\

Conversely, a similar calculation shows that for lower powers of $n$ in $\omega$ the vacuum expectation value diverges (and due to (\ref{proof_1}) also if $p$=0). One should therefore either change the test functions and not use form factors or study different theories. In principle, we could consider free Maxwell electrodynamics, the Proca action or even the free graviton theory and study their behaviour under a renormalisation group flow with the methods of \cite{LLT1}. But here we have altered the Hamiltonian $H$ in order to ensure that the expectation values of holonomies with respect to the vacuum (which is annihilated by $H$) are well-defined. This happens by introducing a higher order polynomial in the Laplacian  (\ref{covariance_s33}) which of course breaks Lorentz invariance. However, our model just serves to test theories with well-defined holonomy operators (but not well-defined electric flux operators) in the usual Fock space setting. Ultimately, we will be interested in coupling general relativity to gauge theories. In this case, theories such as Loop Quantum Gravity \cite{Thi07,AL95,Rov04} indicate that insertion of such  Lorentz invariance breaking higher polynomials is not necessary\cite{QSDV,LT16}.\\

\subsubsection{Initial discretisation on cubic lattice}
In order to test the coarse graining maps on the quantum level, we need to first introduce a discretisation of the phase space of $(E^a(x) ,A_a(y))$ with $a=1,2,3$ with symplectic structure (\ref{sympl_Struc}) and a discretisation of the Hamiltonian (\ref{Hamiltonian_vectorbosons}).\\

We work on a cubic lattice, with $M$ vertices in each direction labelled by $m\in\mathbb{Z}_M^D=\{0,1,...,M-1\}^D$ with $D=3$. At each vertex $m$ we have three in- and three outgoing edges.
We use smearings against form factors to discretise the Hamiltonian. Denoting the edges on the lattice by $e_{m,a}$ (labelled by initial point $m$ and a direction $a=1,2,3$ and $\dot{e}(t)=\hat{a}$ for all $t\in[0,1]$) we have 
\begin{align}
{\rm A}_{M,a}(m)&:=A[F_{e_{m,a}}]=\int_{[0,1]}\; {\rm dt}\; A_a(e_{m,a}(t)),\\
{\rm E}^a_M(m)&:=E[{\rm f}^{S(e_{m,a})}]=\int_{[0,1]^2}\; {\rm d}u\;{\rm d}v\;\; E^a(S_{e_{m,a}}(u,v)),
\end{align}

Similar to section \ref{s23} we interpret this structure as three different field species $a=1,2,3$ (This is due to $D=3$. To make the distinction between directions and field species clear, we will write in this section an arbitrary $D$ for directions but keep $a=1,2,3$ for the field species). Moreover, at each $m$ the field specie $a$ is supported only on edges along direction $\hat{a}$. In order to distinguish the a priori different species, we associate to each of them their own discretised $\omega_{M,a}$, while of course our initial discretisation is such that
\begin{align}\label{initial_same_omega}
\omega_{M,1}=\omega_{M,2}=\omega_{M,3}=\omega_{M}^{(0)}
\end{align}
with $\omega_M^{(0)}$ some discretisation of (\ref{covariance_s33}), such that $\lim_{M\to \infty} \omega_M^{(0)}=\omega$.\footnote{Indeed, we will see in the next sections that the coarse graining induces different flows of $\omega_{M,a}$ for different $a$ in case of the filling kernel, leading ultimately to different fixed pointed families $\{\omega_{M,a}^*\}_M$. However, this ``direction dependence'' is artificial in the sense that it is only present for finite $M$, while in the continuum limit $M\to \infty$ the covariances of all species $a$ agree.}

Since the Hamiltonian is of free harmonic oscillator form for each $a$, we can repeat the discussion from section \ref{s23}:
We introduce the discrete annihilation and creation fields
\begin{align}\label{disc_annihilation_s3}
{\bf a}_{M,a}(m) := \frac{1}{\sqrt{2\hbar}}\big(
\sqrt{\omega_{M,a} \epsilon_M} {\rm A}_{M,a}(m)-i \sqrt{\omega_{M,a} \epsilon_M}^{-1} {\rm E}^{a}_M(m)  \big)
\end{align}
such that
\begin{align}\label{disc_Hamiltonian_s3}
H_M = \hbar \sum_{a=1,2,3} \sum_{m\in\mathbb{Z}^D_M}  \overline{{\bf a}_{M,a}}(m)(\omega_{M,a} {\bf a}_{M,a})(m)
\end{align}
For each specie $a$, we define the corresponding Hilbert spaces $\mathcal{H}_{M,a}$ with Fock vacuum $\Omega_M^{(0)}\in\mathcal{H}_M:=\otimes_a \mathcal{H}_{M,a}$ annihilated by each (\ref{disc_annihilation_s3}) and thus simultaneously by (\ref{disc_Hamiltonian_s3}).
Denoting by $\langle .,.\rangle_{\mathcal {H}_M}$ the scalar product on $\mathcal{H}_M$ it follows (with $f_{M,a}(m)\in L_M:=L(\mathcal{P}^M,\mathbb{R})\equiv\ell_2(\mathbb{Z}_M^{D})^3$)
\begin{align}
\langle \Omega_M, e^{i \widehat{ {\rm A}}_{M}[f_M]} \Omega_M\rangle_{\mathcal{H}_M} = \prod_{a=1,2,3} e^{-\frac{\hbar}{4}\langle f_{M,a},\omega_{M,a}^{-1} f_{M,a}  \rangle} =e^{- \langle f_M,c_M f_M \rangle/2 }
\end{align}
with covariance $c_M ={\rm diag}(c_{M,1},c_{M,2},c_{M,3})$ and $c_{M,a}=\frac{\hbar}{2} \omega_{M,a}^{-1}$. As Gaussian measures do not change their structure under coarse graining \cite{LLT2} the task boils down to find a fixed pointed family $\{c_M^*\}_M$ for the coarse graining maps $J_{M\to2M}$ of both the deleting as well as the filling kernel. Then, we can also use that the fixed pointed Hamiltonian is given by (\ref{disc_Hamiltonian_s3}) when replacing $\omega_{M,a} \mapsto \omega_{M,a}^*$.\\

Also, we discussed already in section \ref{s23} that the fixed point for choice $n\geq4$ in (\ref{covariance_s33}) can be achieved by finding the fixed point of $n=2$ due to the fact that both are related via the contour integral (\ref{inverseresidualtheorem}) and replacing $p^2\mapsto \bar{p}^2=p^2-k_o$.\\

We end this section by choosing an explicit initial discretisation of the covariance, i.e. $c_M^{(0)}={\rm diag}(c_{M,1}^{(0)},c_{M,2}^{(0)},c_{M,3}^{(0)})$, which acts on test functions $f_{M,a}(m)\in L_M^3$. We assume that every field specie has a translational invariant covariance, i.e. its kernel is for $m,n\in\mathbb{Z}^D_M$:
\begin{align}
c_{M,a}(m,n) =c_{M,a}(m-n)
\end{align}
which holds true for the following initial discretisation of the derivatives inside 
\begin{align}
\omega^2_M=(-\sum_{b=1}^D \partial_b^M \partial^b_M+\bar{p}^2)
\end{align}
with:
\begin{align}
(\partial_b^M f_M)_a(m):=\frac{1}{\epsilon_M}[f_{M,a}(m)-f_{M,a}(m-\hat{b})]\\
(\partial^b_M f_M)_a(m):=\frac{1}{\epsilon_M}[f_{M,a}(m)-f_{M,a}(m+\hat{b})]
\end{align}
and $\hat{b}$ is the normal vector pointing in direction $b$. We see that $c_M$ does not mix the different species $a$, therefore we can apply the discrete Fourier transform from (\ref{disc_FT}) on each subspace of fixed $a$ to get as initial starting point for the covariance (see \cite{LLT2} for details):
\begin{align}\label{initial_covariance_s3}
\hat{c}_{M,a}^{(0)}(l) = \frac{\hbar/2}{\bar{p}^2-\epsilon_M^{-2}\sum_{d=1}^{D}(2\cos(k_Ml_d)-2)}
\end{align}
with $l=\{l_1,...l_D\}$ and $k_M=2\pi/M$. Note that the right-hand side of (\ref{initial_covariance_s3}) is independent of $a$ due to the initial choice (\ref{initial_same_omega}). This will change once we study the RG flow of the filling kernel.\\
Lastly, let us recall from \cite {LLT4} that an initial covariance of the form (\ref{initial_covariance_s3}) can be transformed via the residue theorem into several integrals over a product of ``one-dimensional'' covariances, i.e. decouples in each direction:
\begin{align}\label{facotrisation}
\hat{c}_{M,a}^{(0)}(l) = \frac{\hbar/2}{(4\pi i)^3}(\prod_{b=1}^D\oint_\Gamma {\rm d}z_b) \; \frac{1}{\bar{p}^2-\sum_b z_b}\;c_M'(l_a,a;z_a)\; \prod_{b\neq a}  c'_M(l_{b},a;z_b)
\end{align}
where $\Gamma$ is a contour surrounding the real axes (closing at $\pm\infty$ and thus including both poles) and  ($l'\in\mathbb{Z}_M$)
\begin{align}\label{initial_1dim_cov_for_s3}
c'_M(l',a;z) :=\frac{1}{z-\epsilon_M^{-2}(2\cos(k_M l')-2)}
\end{align}
Note that the way in which split the integrals is purely conventional and does not affect the continuum limit $M\to \infty$. Also, the initial covariance does not have a direction dependency, hence the label $a$ does not appear on the right hand side of (\ref{initial_1dim_cov_for_s3}).\\
A factorisation property like (\ref{facotrisation}) becomes useful if it can be established that the covariance does not change this structure under a renormalisation step. In such a case, each of the $c'_M$ will drive into its respective fixed point \cite{LLT4}. Indeed, this will be case for both the deleting and the filling kernel as we discuss in the next two sections. There, we will study the different Hamiltonian RG flows in order to find the fixed pointed covariances $c_{M,a}^*=\frac{\hbar}{2}{(\omega_{M,a}^*)}^{-1}$ for each field specie $a$. They completely describe the Hilbert spaces and the Hamiltonians at finite resolution.

\subsection{Toy model: Fixed points of the deleting kernel}
\label{s34_del}
From now on we set $D=3$ explicitly in all formulae. We study the RG flow of the coarse graining for the deleting kernel from (\ref{quantum_coarse_graining}), i.e.
\begin{align}
J_{M\to 2M} (e^{i \widehat{{\rm A}}_{M}[f_M]}\Omega_M^{(n+1)}):= e^{i\widehat {{\rm A}}_{2M}[I^{Del}_{M\to2M}f_M]}\Omega_{2M}^{(n)}
\end{align}
which is equivalent to the flow of the family of Hilbert space measures 
\begin{align}
\langle \Omega_M^{(n+1)}, e^{i \widehat{\rm A}_M [f_M]}\Omega_M^{(n+1)}\rangle_{\mathcal{H}_M^{(n+1)}}= \langle \Omega_{2M}^{(n)},e^{i \widehat{\rm A}_{2M}[I_{M\to 2M}^{Del}f_M]}\Omega_{2M}^{(n)}\rangle_{\mathcal{H}_{2M}^{(n)}}
\end{align}
that is (see \cite{LLT2}):
\begin{align}\label{RG_flow_covariance}
\langle f_M , c^{(n+1)}_M g_M\rangle_M:= \langle I_{M\to 2M}^{Del}\; f_M, c^{(n)}_{2M}\; I^{Del}_{M\to 2M}\; g_M\rangle_{2M}
\end{align}
with $f_M,g_M\in L_M =\ell_2(\mathbb{Z}^3_M)^3$ and deleting kernel ($m'\in\mathbb{Z}^3_{2M}$)
\begin{align}\label{deleting_explicit}
(I_{M\to 2M}^{Del}f_M)_a(m')=\sum_{m\in\mathbb{Z}^3_M} f_{M,a}(m) (\delta_{m',2m}+\delta_{m',2m+\hat{a}})
\end{align}
We see that (\ref{deleting_explicit}) does not mix the field species for different $a$ with each other and does not distinguish between different $a$. Together with the fact that the initial covariance was written as diagonal matrix $c_M={\rm diag}(c_{M,1},c_{M,2},c_{M,3})$, this implies that the same holds at each iteration of the RG flow and thus also the fixed point measure will be a product of three times the same Gaussian measures for each $a$.\\
However, for each field specie $a$ the different directions with respect to the lattices vertices $m=(m_1,m_2,m_3)$ behave differently as $\hat{a}$ enters the right-hand side of (\ref{deleting_explicit}). Thus, in direction $\hat{a}$ the $I_{M\to2M}^{Del}$ behaves as the one-dimensional blocking kernel studied in \cite{LLT2}, that is 
\begin{align}
(I_{M\to2M}^{Del} f_M)_a(2m_a,...)=(I_{M\to2M}^{Del} f_M)_a(2m_a+1,...)
\end{align}
for $m_a\in\mathbb{Z}_M$ being the $a^{\rm th}$ component of $m\in\mathbb{Z}_M^3$. However, for $b\neq a$, the kernel behaves as the one-dimensional deleting kernel from \cite{LLT3}, that is for $m_b$ being the $b^{\rm th}$ component of $m$:
\begin{align}
(I_{M\to2M}^{Del} f_M)_a(...,2m_b+1)=0
\end{align}
Thus, the flow of the coarse graining map from (\ref{deleting_explicit}) introduces a ``direction dependence'' of the covariance at the quantum level for finite resolution $M$. This dependence only vanishes in the continuum limit $M\to\infty$.\\
Since the RG flow in (\ref{RG_flow_covariance}) does not mix the different directions, for a decoupled covariance of the form (\ref{facotrisation}) each ``one-dimensional covariance'' $c'_M$ will flow into its respective fixed point. And since the RG flows for direction $a$ and $b\neq a$ behave like the ones of the injections studied in \cite{LLT2} and \cite{LLT3} respectively, the fixed points are already known and read:
\begin{align}
(\hat{c}'_M)^*(l',b;z)=\hat{c}_M^{Del} (l',z)   &\hspace{60pt}\text{ for direction } b\neq a\\
(\hat{c}'_M)^*(l',a;z)=\hat{\underline{c}}_M (l',z)\hspace{4.5pt} &\hspace{60pt}\text{ for\;direction\;} a
\end{align}
with $l'\in\mathbb{Z}_M,z\in\mathbb{C}$ and the definitions from (\ref{standard_blocking_fp}) and (\ref{standard_del_fp}).\\
It remains to plug the fixed points for each direction into (\ref{facotrisation}) and to restore the correct $n$-dependence via (\ref{inverseresidualtheorem}). Thus, we know the complete fixed pointed covariance $c^*={\rm diag}(c_{M,1}^*,c_{M,2}^*,c_{M,3}^*)$ with the following kernels for the Fourier transform of the covariances:
\begin{align}\label{final_result_deleting_kernel}
\hat{c}^*_{M,a}(l)=\frac{p^{(n-1)}\hbar}{(4\pi i)^4}\oint_\gamma {\rm d}k_o\;\frac{1}{k_o^{n/2}}\;(\prod_{b=1}^3 \;\;\oint_\Gamma {\rm d}z_b)\;\;\frac{1}{\bar{p}^2-\sum_{b=1}^3 z_b}\; \hat{\underline c}_M(l_a,z_a)\prod_{b\neq a}\;\; \hat{c}^{Del}_M(l_b, z_b)
\end{align}
where we remember that $\bar{p}^2=p^2-k_o$. For further details, see \cite{LLT2,LLT3,LLT4}.

\subsection{Toy model: Fixed points of the filling kernel}
\label{s35_fil}
We turn towards the second choice of coarse graining map that was motivated in this paper. While of course further coarse graining maps can be constructed, the analysis of this section presents already an indication of universality -- as it will transpire that the continuum limit $M\to\infty$ of the fixed pointed theories for both kernels agree.\\
Again it is $D$=3. The three different choices of filling kernels are labelled by ${\rm i}\in\{1,2,3\}$ and their explicit action is obtained by using the form of the characteristic functions in (\ref{parallelogramlattice}):
\begin{align}\label{filling_kernel_explicit}
&(I^{\rm Fil,i}_{M\to 2M}f_M)_a(m')=\\
&=\sum_{m\in\mathbb{Z}^3_M}   f_{M,a}(m)\begin{cases}  \delta_{\lfloor(m_{a}'-\sum_{b\neq a}\delta^{odd}(m_b'))/2\rfloor,m_{a}}\prod_{b\neq a}\delta_{\lfloor m_b'/2\rfloor,m_b}& \text{for } {\rm i}=a\\
\delta_{\lfloor(m'_a-\delta^{odd}(m_{\rm i}')+\delta^{odd}(m_s'))/2\rfloor,m_a}\delta_{\lfloor( m_{\rm i}'+\delta^{odd}(m'_s) )/2\rfloor,m_{\rm i}}\delta_{\lfloor m'_s/2\rfloor,m_s}& \text{else}
\end{cases}\nonumber
\end{align}
where $s=1,2,3$ is distinct from both $a,{\rm i}$, that is $a\neq s\neq {\rm i}$ and $\delta^{odd}(n)=0$ iff $n\in 2\mathbb{Z}_M$ and $=1$ else. Like in the previous subsection, we see that different field species $a$ will not talk to each other, therefore keeping the structure $c_M={\rm diag}(c_{M,1},c_{M,2},c_{M,3})$ intact during the whole RG flow.\\
However, a notable difference to the map $I_{M\to2M}^{Del}$ is that the choice of ${\rm i}$ leads to different fixed pointed families for the field specie labelled by $a$ -- since (\ref{filling_kernel_explicit}) singles out the case with $a={\rm i}$. On top of that, the directions of the lattice vertices $m\in\mathbb{Z}^3_M$ do {\it not} decouple in an obvious way. Thus, we need to carefully study how the matrix elements of a covariance transform under a renormalisation step, which reads for fixed $a$ and $f_M,g_M\in L_M$
\begin{align}
\langle f_{M,a} , c^{(n+1)}_{M,a} g_{M,a}\rangle_M:= \langle I_{M\to 2M}^{\rm Fil,i} f_{M,a} , c^{(n)}_{2M,a} I^{\rm Fil,i}_{M\to 2M} g_{M,a}\rangle_{2M}
\end{align}
Here, we only show the case $a={\rm i}=1$ explicitly, all other choices work analogously. By writing explicitly $\langle .,.\rangle_M$ with (\ref{scalarproduct_testfunctions}) abbreviating $f_M(m):=f_{M,a=1}(m)$ and plugging in (\ref{filling_kernel_explicit}), we perform the following manipulations:
\begin{align}
&\epsilon^{-2D}_{2M}\langle f_M , c^{(n+1)}_{M,1} g_M\rangle_M=\sum_{m,\tilde{m}\in \mathbb{Z}^3_{2M}}(I_{M\to2M}^{\rm Fil,1}f_M)(m)(I_{M\to 2M}^{\rm Fil,1}g_M)(\tilde{m})c^{(n)}_{2M,1}(m,\tilde{m})=\\
&=\sum_{m_1',m'_2,m_3'\in\mathbb{Z}_{2M}}\sum_{\tilde{m}_1',\tilde{m}_2',\tilde{m}_3'\in\mathbb{Z}_{2M}}c^{(n)}_{2M,1}(m_1',m_2',m_3',\tilde{m}_1',\tilde{m}_2',\tilde{m}_3')\;\times\\
&\;\;\;\times\; f_M(\lfloor \frac{m'_1-\delta^{odd}(m'_2)-\delta^{odd}(m'_3)}{2} \rfloor,\lfloor\frac{m_2'}{2} \rfloor,\lfloor \frac{m_3'}{2} \rfloor)\; g_M(\lfloor \frac{\tilde{m}'_1-\delta^{odd}(\tilde{m}'_2)-\delta^{odd}(\tilde{m}'_3)}{2} \rfloor,\lfloor\frac{\tilde{m}_2'}{2} \rfloor,\lfloor \frac{\tilde{m}_3'}{2} \rfloor)\nonumber\\
&=\sum_{m,\tilde{m}\in\mathbb{Z}_M^3}\sum_{\delta,\tilde{\delta}\in\{0,1\}^3}c^{(n)}_{2M,1}(2m_1+\delta_1,2m_2+\delta_2,2m_3+\delta_3,2\tilde{m}_1+\delta_1,2\tilde{m}_2+\delta_2,2\tilde{m}_3+\delta_3)\;\times\nonumber\\
&\;\;\;\times\;f_M(m_1+\lfloor(\delta_1-\delta_2-\delta_3)/2\rfloor,m_2,m_3)g_M(\tilde{m}_1+\lfloor(\tilde{\delta}_1-\tilde{\delta}_2-\tilde{\delta}_3)/2\rfloor,\tilde{m}_2,\tilde{m}_3)
\end{align}
where in the last step we expressed $m'_i=2 m_i + \delta_i$ with $m,\tilde{m}\in\mathbb{Z}_M ^3$ and $\delta,\tilde{\delta}\in\{0,1\}^3$. We  can now shift the summation parameter $m_1 \mapsto m_1-\lfloor( \delta_1-\delta_2-\delta_3)/2\rfloor$ using that $f_M(0)=f_M(M)$ and $c^ {(n)}_{2M,1}(0,...)=c^{(n)}_{2M,1}(2M,....)$:
\begin{align}
&\epsilon^{-2D}_{2M}\langle f_M , c^{(n+1)}_{M,1} g_M\rangle_M=\sum_{m,\tilde{m}\in\mathbb{Z}_M^3}f_M(m)g_M(m) \times\\
&\;\;\;\times\;\sum_{\delta,\tilde{\delta}\in\mathbb{Z}^D_M} c^{(n)}_{2M,1}(2m_1+\delta_1-2\lfloor\frac{\delta_1-\delta_2-\delta_3}{2}\rfloor,...,2\tilde{m}_1+\tilde{\delta}_1-2\lfloor\frac{\tilde{\delta}_1-\tilde{\delta}_2-\tilde{\delta}_3}{2}\rfloor,2\tilde{m}_2+\tilde{\delta}_2,...)\nonumber
\end{align}
As this equation is for arbitrary $f_M,g_M$, it must hold component wise and gives us the following recursion relation for the RG flow: 
\begin{align}
c^{(n+1)}_{M,1}(m,\tilde{m})=2^{-2D}\sum_{\delta,\tilde{\delta}\in\mathbb{Z}^3_M}c^{(n)}_{2M,1}(2m_1+\sum_i\delta_i,2m_2+\delta_2,2m_3+\delta_3,2\tilde{m}_1+\sum_i\tilde{\delta}_i,2\tilde m_2+\tilde\delta_2,2\tilde {m}_3+\tilde{\delta}_3)\nonumber
\end{align}
where we realized that $\sum_i \delta_i$ can be obtained by interchanging the summation parameter $\delta_i\mapsto  \delta_i+1$ in the cases where $\delta_2+\delta_3=1$.\\
In order to proceed, we employ the assumption of the covariance to be translational invariant, i.e. $c_{2M,1}(m,\tilde{m})=c_{2M,1}(m-\tilde{m})$, and go into Fourier space, where the recursion relation reads:
\begin{align}
&\hat{c}^{(n+1)}_{M,1}(l)=2^{-2D}\sum_{\delta,\tilde{\delta},\eta\in\{0,1\}^3} \hat{c}^{(n)}_{2M,1}(l+\eta M)\exp(ik_{2M}(l+\eta M)\cdot [\left(\begin{array}{c}
\sum_i\delta_i\\
\delta_2\\
\delta_3
\end{array}\right)-\left(\begin{array}{c}
\sum_i\tilde{\delta}_i\\
\tilde{\delta}_2\\
\tilde{\delta}_3
\end{array}\right)])\nonumber\\
&=2^{-2D}\sum_{\eta\in\{0,1\}^3}\hat{c}^{(n)}_{2M,1}(l+\eta M)[1+e^{i k_{2M}(l_1+\eta_1 M)}][1+e^{ik_{2M}(l_1+l_2+(\eta_1+\eta_2)M)}][1+e^{ik_{2M}(l_1+l_3+(\eta_1+\eta_3)M)}]\nonumber\\
&\;\;\;\;\times [1+e^{-i k_{2M}(l_1+\eta_1 M)}][1+e^{-ik_{2M}(l_1+l_2+(\eta_1+\eta_2)M)}][1+e^{-ik_{2M}(l_1+l_3+(\eta_1+\eta_3)M)}]\label{filling_kernel_recursion}\\
&=2^{-2D}\sum_{\eta\in\{0,1\}^3}\hat{c}^{(n)}_{2M,1}(l+\eta M)2^D[1+\cos(k_{2M}(l_1+\eta_1 M))]\times\nonumber\\
&\hspace{50pt}\times[1+\cos(k_{2M}(l_1+l_2+(\eta_1+\eta_2) M))][1+\cos(k_{2M}(l_1+l_3+(\eta_1+\eta_3) M))]\nonumber\\
&=2^{-D}\sum_{\delta\in\{0,1\}^3}{\hat{\bf c}_{2M,1}}^{(n)}(l_1+\delta_1 M,(l_1+l_2)+\delta_2 M,(l_1+l_3)+\delta_3 M) \times\nonumber\\
&\hspace{50pt}\times[1+(-)^{\delta_1}\cos(l_1)][1+(-)^{\delta_2}\cos(k_{2M}(l_1+l_2))][1+(-)^{\delta_3}\cos(k_{2M}(l_1+l_3)]\nonumber
\end{align}
where in the last step we introduced $\hat{\bf c}_{2M,1}(l_1,l_1+l_2,l_1+l_3)\equiv\hat{c}_{2M,1}(l_1,(l_1+l_2)-l_1,(l_1+l_3)-l_1)$ and used the periodicity $\hat{\bf c}_{2M,a}(l+2M)=\hat{\bf c}_{2M,a}(l)$ to relabel $\eta\to \delta$ and $\cos(l+k_{2M} M)=-\cos(l)$ (due to $k_{2M}M=\pi$) .\\
We observe that if the initial covariance could be written as a product of the form $c'(l_1)c'(l_1+l_2)c'(l_1+l_3)$ then every element of the RG flow would have this property (similar to \cite{LLT4}). Thus, we aim at splitting $\hat{c}^{(0)}_M$ via another application of (\ref{facotrisation}). For this purpose, note the following identity for $\hat{c}^{(0)}_M$ from (\ref{initial_covariance_s3}):
\begin{align}\label{splitting_s3}
\hat{c}^{(0)}_{M,a}(l)=\frac{\hbar}{2}\Xi_a^{-1}[z_1,z_2,z_3]\mid_{z_a=\epsilon_M^{-2}(2\cos(k_Ml_a)-2),\; z_{b\neq a}=\epsilon^{-2}_M(2\cos(k_M(l_a+l_b))-2)}
\end{align}
with 
\begin{align}
\Xi_a[\{z\}]:=\bar{p}^2-z_a-2\epsilon^{-2}_M\sum_{b\neq a}\left[
(\frac{\epsilon_M^2}{2}z_a+1)(\frac{\epsilon_M^2}{2}z_b+1)+\sqrt{1-(\frac{\epsilon_M^2}{2}z_a+1)^2}\sqrt{1-(\frac{\epsilon_M^2}{2}z_b+1)^2}
-1\right]\nonumber
\end{align}
Hence, with (\ref{facotrisation}) and $c'_M$ from (\ref{initial_1dim_cov_for_s3}) we find the desired splitting:
\begin{align}
\hat{c}^{(0)}_{M,a}(l)=\frac{\hbar/2}{(4\pi i)^3}\;(\prod_{b=1}^3 \;\;\oint_\Gamma {\rm d}z_b)\; \Xi^{-1}[\{z\}]\; c'_M(l_a,a;z_a)\prod_{b\neq a} c'_M(l_a+l_b,a;z_b)
\end{align}
Moreover, the recursion with $[1\pm\cos(...)]$ is the same as in \cite{LLT2} and thus is known to lead to the fixed point $\underline{\hat{c}}_M$ from (\ref{standard_blocking_fp}). In other words, we know to which fixed point family the flow induced by recursion (\ref{filling_kernel_recursion}) drives to.
Lastly, we again restore the contour integral to take $n\geq 4$ into account and obtain the final result:
\begin{align}\label{final_fixed_point}
\hat{c}^*_{M,a={\rm i}}(l)=\frac{p^{(n-1)}\hbar}{(4\pi i)^3}\oint_\gamma \frac{{\rm d}k_o}{k_o^{n/2}}\;(\prod_{b=1}^3\oint_\Gamma {\rm d}z_b)\;\Xi_a[\{z\}]\;\;\hat{\underline{c}}_M(z_a,l_a)\; \prod_{b\neq a} \hat{\underline{c}}_M(z_b,l_a+l_b) 
\end{align}
Analogously, iterating the same steps for $a\neq {\rm i}$ we get: (with $a\neq s\neq {\rm i}$)
\begin{align}
\hat{c}^*_{M,a\neq{\rm i}}(l)=\frac{p^{(n-1)}\hbar}{(4\pi i)^3}\oint_\gamma \frac{{\rm d}k_o}{k_o^{n/2}}\;(\prod_{b=1}^3\oint_\Gamma {\rm d}z_b)\;\Xi_a'[\{z\}]\;\;\hat{\underline{c}}_M(z_a,l_a)\; \hat{\underline{c}}_M(z_{\rm i},l_a+l_{\rm i})\; \hat{\underline{c}}_M(z_s,l_{\rm i}+l_s-l_a) \nonumber
\end{align}
where $\Xi'_a$ can be obtained from a similar splitting as in (\ref{splitting_s3}).\\

If one performs the continuum limit $M\to \infty$, one sees that the artificial direction dependence as well as the difference between the field species $a$ will be lost and the continuum theory agrees thus with the continuum limit from (\ref{final_result_deleting_kernel}), i.e. the fixed point of the deleting kernel. In other words, the projections of same continuum theory with different coarse graining projections carrying the same label $M$ differ. Yet, the difference is merely due to the fact that the coarse graining maps are different. The continuum theory is in both cases the same and thus displays universality with respect to change of the coarse graining map.\\

\section{Conclusion}
\label{s4_conclusion}

In this paper we performed preliminary steps to extend the Hamiltonian Renormalisation Group to Abelian gauge theories. This  serves as a further step towards the construction of interacting QFTs for those systems which are subject to constraints.\\
When constraints are present, a possible strategy is to perform a symplectic reduction and go to the reduced phase space on which the constraints have been implemented. In general, the geometry (i.e. the symplectic structure) of the reduced phase space may be very complicated, but at least for the Gauss constraint of Abelian gauge theories the procedure is well understood: one can split the phase space in transversal and longitudinal modes and then gauge-fix the unphysical longitudinal modes. This allows to proceed with canonical quantisation and renormalisation along the methods for scalar fields from \cite{LLT1,LLT2,LLT3,LLT4}. In a class of models that includes free Maxwell theory we performed a reduced phase space quantisation obtaining a family of Fock Hilbert spaces $\mathcal{H}_M$, one for each resolution $M$. For this class, we could test different injections $J_{M\to 2M}:\mathcal H_M \to \mathcal{H}_{2M}$. It transpired that the resulting models can be understood as two decoupled field species, both of them running into their fixed point, which we knew analytically due to previous studies in \cite{LLT2,LLT3}.\\
The reduced phase space approach results in  a renormalisation flow which  is very close to that of scalar fields. In order to test renormalisation flows that take the vector field structure into account we considered a second class of models without Gauss constraint which includes free Proca theory. The motivation for considering generalisations of free Maxwell and Proca theory is that some of these models allow for well defined holonomy operators in the corresponding Fock representations at the price of losing Poincar\'e invariance. We consider these models as mere toy models for quantum gravity theories \cite{QSDV,LT16} that are based on Hilbert space representations with both well defined holonomy operators and Hamiltonians without breaking symmetries. In particular we are thinking about discretisations of the Hamiltonian operators studied in this paper using holonomies themselves which would simulate the proposal of \cite{QSDV,LT16}. In a future publication \cite{tbp} we will also aim at imposing the Gauss constraint after quantisation.
The idea of introducing a ``smoothening'' operator into  the Hamiltonian in order to allow for holonomy operators in the corresponding Fock  representation is in some sense dual to the idea of using  smoothened form factors studied in \cite{Var99}. Note also that we could have made our deformation of Proca or Maxwell theory phenomenologically more interesting by changing $\omega^2 \mapsto -\Delta+p^2 ((-\Delta+p^2)/\mu^2)^n$ with $p$ arbitrarily small but finite and $\mu $ arbitrarily large but  finite so that the Lorentz violation will only manifest itself at energies above $\mu$. Even in this case holonomies are still well-defined operators and the presented strategy to determine the fixed point remains the same.\\
We chose two different coarse graining maps in order to understand how stable the fixed points of the theory are under changes of the injection maps. Both maps -- deleting and filling kernel -- are mathematical well-defined, but the level of experience that we have for them differs: the deleting kernel has already been actively studied in the literature and found application in the non-Abelian case of Loop Quantum Gravity where it enabled the construction of an inductive limit Hilbert space. Spin networks (a possible basis of said Hilbert space) carry distributional excitations such that a smooth quantum geometry can only be obtained by distributions on the Hilbert space. Conceptually, reobtaining smooth geometry could be easier when working with the filling kernel, as it excites all edges as the resolution increases. However, extensive studies on the latter kernel have not been performed as of today.\\
Both maps employ discretisations of the spatial manifold where the fields are smeared along edges of a cuboidal lattice. Choosing such cubic lattices might at first glance look like a restriction of the theory since it gives rise to the so-called ``staircase problem'' \cite{STW01}: albeit square lattices suffice to separate the points in phase space as $M$ gets large, one does not have access to ``45'' degree line observables at any finite resolution. Yet, the continuum theory does allow considering holonomy operators along such curves which are not straight. This stresses the point that the lattice just serves to construct the continuum theories, all other investigations have to start from there.\\
We demonstrated for our model classes that the relevant fixed points can be found for the filling as well as for the deleting  kernel. Due to the fact that the discretisations were expressed in terms of smearings with form factors, the investigation exploited many of the findings from previous applications of the Hamiltonian Renormalisation Group in \cite{LLT2,LLT3,LLT4}. Finally, we found analytically closed formulas for the respective fixed points and saw that the Hamiltonian renormalisation leads to reliable results. 

\section*{Acknowledgements}
The authors thank Thorsten Lang for many fruitful discussion. This work was funded by the project BA 4966/1-2 of the German Research
Foundation (DFG). KL also acknowledges support by the DFG under Germany’s
Excellence Strategy – EXC 2121 ``Quantum Universe'' – 390833306.

}

\end{document}